# PULSE: Palomar Ultraviolet Laser for the Study of Exoplanets

**Christoph Baranec,**
*Institute for Astronomy, University of Hawai'i at Manoa, Hilo, HI 96720*
**Richard Dekany,**
*Caltech Optical Observatories, California Institute of Technology, Pasadena, CA 91125*
**Marcos van Dam,**
*Flat Wavefronts, Christchurch, 8140, New Zealand*
**& Rick Burruss**
*Jet Propulsion Laboratory, California Institute of Technology, Pasadena, CA 91109*

## ABSTRACT

PULSE is a new concept to augment the currently operating 5.1-m Hale PALM-3000 exoplanet adaptive optics system with an ultraviolet Rayleigh laser and associated wavefront sensor. By using an ultraviolet laser to measure the high spatial and temporal order turbulence near the telescope aperture, where it dominates, one can extend the faintness limit of natural guide stars needed by PALM-3000. Initial simulations indicate that very-high infrared contrast ratios and good visible-light adaptive optics performance will be achieved by such an upgraded system on stars as faint as $mV = 16\text{-}17$ using an optimized low-order NGS sensor. This will enable direct imaging searches for, and subsequent characterization of, companions around cool, low-mass stars for the first time, as well as routine visible-light imaging twice as sharp as HST for fainter targets. PULSE will reuse the laser and wavefront sensor technologies developed for the automated Robo-AO laser system currently operating at the Palomar 60-inch telescope, as well as take advantage of pending optimization of low-order NGS wavefront sensing and planned new interfaces to the PALM-3000 real-time reconstruction computer. A copy of the Robo-AO laser will be installed in the prime focus cage of the 5.1-m, and a new ultraviolet high-order wavefront sensor, fed by an ultraviolet dichroic, will be installed in the space above the PALM-3000 optical bench near the calibration sources. The laser measurements will drive the 3,388 active element high-order deformable mirror in open-loop, while an adaptive optics sharpened faint natural source will be measured by the current PALM-3000 wavefront sensor in its lowest spatial sampling mode, with commands sent in closed-loop to the 241 active element low-order deformable mirror. The natural guide star loop corrects for both the relatively weak low-order high-altitude turbulence as well as functioning as both the tip-tilt and low-bandwidth `truth' sensor loops in a traditional laser adaptive optics system.

## 1. INTRODUCTION

A clearer understanding of formation and evolution history of the Earth can be gained through the comparative study of the architecture and physical properties of planetary systems now being discovered in orbit around other stars in our galaxy. Recently commissioned and upcoming astronomical instruments such as PALM-3000 + P1640 [1,2], SCExAO [3], Gemini Planet Imager [4], and SPHERE [5], are opening the a window for direct imaging and spectroscopic observations of exoplanet systems previously obscured by the intense relative brightness of host stars with respect to that light reaching us from hosted exoplanets. These systems rely on the coordinated interworking of sensitive integral field spectrographs, optimized coronagraphs, ultra-accurate wavefront calibration measurements, and atmospheric aberration correction – moving from seeing-limited to diffraction-limited image quality – to suppress scattered starlight to a level sufficient for direct exoplanet studies.

One components of this multi-tiered technology, the atmospheric turbulence compensating adaptive optics (AO) subsystem, currently limits the faintness of stellar systems that may be interrogated directly for exoplanets. The requirement to have available a sufficiently bright natural guide star (NGS) for precision wavefront sensing and compensation limits the utility of the current generation of AO instruments to only stars having an apparent magnitude at visible wavelengths, $mV < 11$. Thus, though they are beginning to reveal fascinating new exoplanet spectra [2], current surveys of exoplanetary systems [6] are constrained to relatively bright stars.

The general constraint on AO observation of appropriate guide star availability has previously motivated the development of strategies for the deployment of synthetic guide star beacons based upon any of several high-power laser technologies. Laser guide star (LGS) adaptive optics has been successfully deployed on many large telescopes since the 1980's to dramatically increase the available fraction of sky for diffraction-limited AO observations.

The Palomar Ultraviolet Laser for the Study of Exoplanets (PULSE) is an extension of the PALM-3000 AO system recently commissioned on the 5-meter diameter Hale telescope at Palomar Observatory [1]. Motivated by our scientific desire to investigate fainter, cooler, and more distant stellar systems in our galactic neighborhood, PULSE will uniquely enable accurate wavefront correction with guide star operation to mV < 16. Specifically, PULSE will for the first time obtain spectra of exoplanets orbiting cool, low-mass stellar hosts, such as M dwarfs, T dwarfs, and cooler spectral types.

Although primarily motivated by the need for this unique exoplanet observing capability, we note that a very wide variety of extragalactic and faint-object science is also enabled by PULSE, utilizing the six backend instruments that are fed by PALM-3000. PULSE will enable diffraction-limiting science at modest infrared Strehl ratios utilizing guide stars as faint as mV < 18, making virtually the entire night sky from Palomar available for high angular resolution study.

## 2. EXAMPLE OF CURRENT HIGH-CONTRAST ADAPTIVE OPTICS

PALM-3000 is a second generation astronomical adaptive optics facility instrument for the 5.1 meter Hale telescope at Palomar Observatory [1]. PALM-3000 uses two Xinetics, Inc. deformable mirrors in series as a woofer-tweeter pair to correct for measured atmospheric conditions at Palomar Mountain. The low-spatial frequency, large stroke mirror is the original PALM-241 [7], which has been in use since 1999. The high-spatial frequency, small stroke mirror is a new 3388 actuator device, the largest format astronomical deformable mirror used on the sky to date [8].

PALM-3000 uses a Shack-Hartmann wavefront sensor with four selectable pupil sampling modes [9] to measure the incoming wavefront. The detector is a 128x128 pixel E2V CCD50 encased in a SciMeasure camera head, originally developed for PALM-241 [10]. Pupil sampling modes of 64x, 32x, and 8x subapertures across the entrance pupil allow for performance optimizations on guide stars spanning 18 magnitudes of brightness [11].

Real-time computation is performed by custom software operating on 16 graphics cards distributed over 8 computers, all connected to a servo control and database computer via a high speed network switch [12]. Wavefront reconstruction is optimized on each target by a reconstructor pipeline tool, which automatically gathers wavefront sensor pixel data and weights each controlled actuator based on the flux measured in its matched subaperture. The PALM-300 operator may optimize the wavefront further by adjusting servo control gains while monitoring telemetry diagnostic displays.

Five science instruments have been commissioned with PALM-3000: PHARO, a 1k x 1k HgCdTe imager, grism spectrograph, and coronagraph [13]; PFN, a fiber nulling interferometer [14]; SWIFT, a 44 x 89 spaxial integral field spectrograph [15]; P1640, a 200 x 200 integral field spectrograph coronagraph with nanometer level metrology [16]; and TMAS, a 2560 x 2560 sCMOS imager [1].

PALM-3000 was commissioned on sky in the summer of 2011 in the 64x pupil sampling mode, and several updates have been implemented since then to improve system performance, stability, and reliability. To date (and in average sky conditions at Palomar Observatory) PALM-3000 has achieved an average of 82% K-band Strehl ratios (150 nm rms) with the PHARO science camera on bright guide stars (mv 3-7), with a best-case image of 85.5% (135 nm rms). PALM-3000 is performing reasonably close to expectations in 64x and 32x modes on stars in the 8-13 mv range, as can be seen in Figure 1. More updates to PALM-3000 are planned through 2014 in an effort to boost bright guide star performance in 64x mode, as well as faint guide star performance in 32x and 8x modes.

Several PALM-3000 science results of note have been recently released, or are now in print. The P1640 coronagraph team obtained spectra of all four known planets orbiting the star HR8799, the first simultaneous spectroscopic observations of multiple exoplanets [2]. In addition, the P1640 team re-characterized a suspected planetary companion to a well-known bright star, reclassifying the object as a brown dwarf [17].

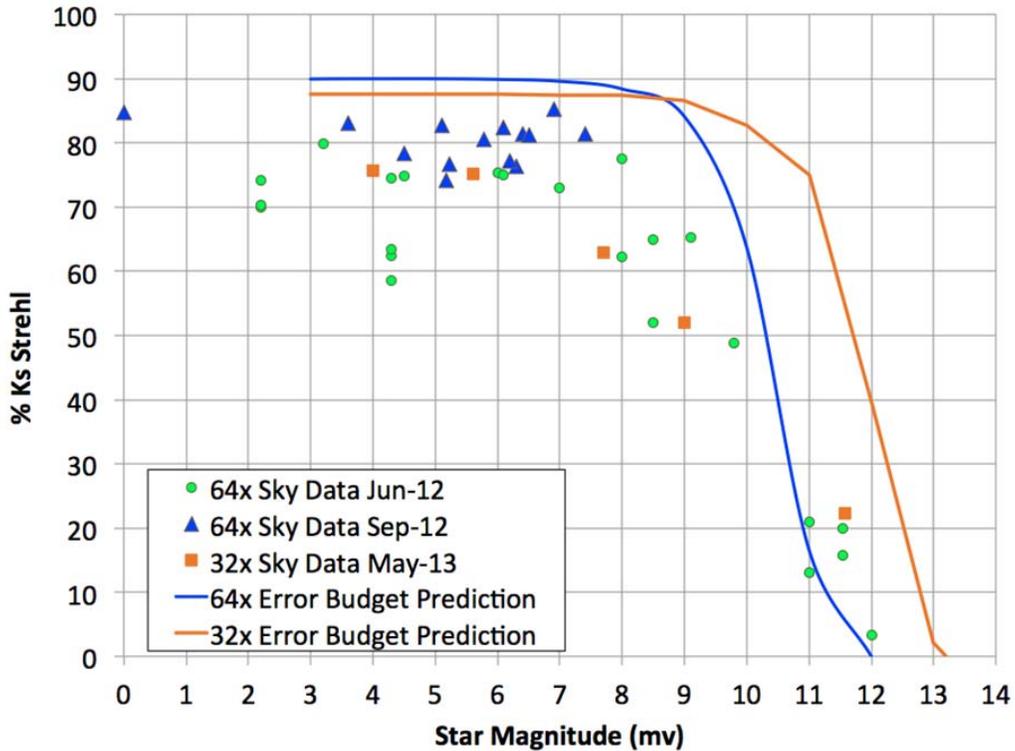

Fig. 1. PALM-3000 on-sky performance to date as a function of guide star magnitude and pupil sampling mode. The performance metric used is $K_{short}$-band Strehl ratio as obtained with the PHARO science camera. Error budget predictions under similar seeing and other observing conditions are represented by the solid curves.

Results in optical wavelengths have also been obtained with PALM-3000, most notably with the SWIFT integral field spectrograph [18] and with the TMAS imaging camera as shown in Figure 2.

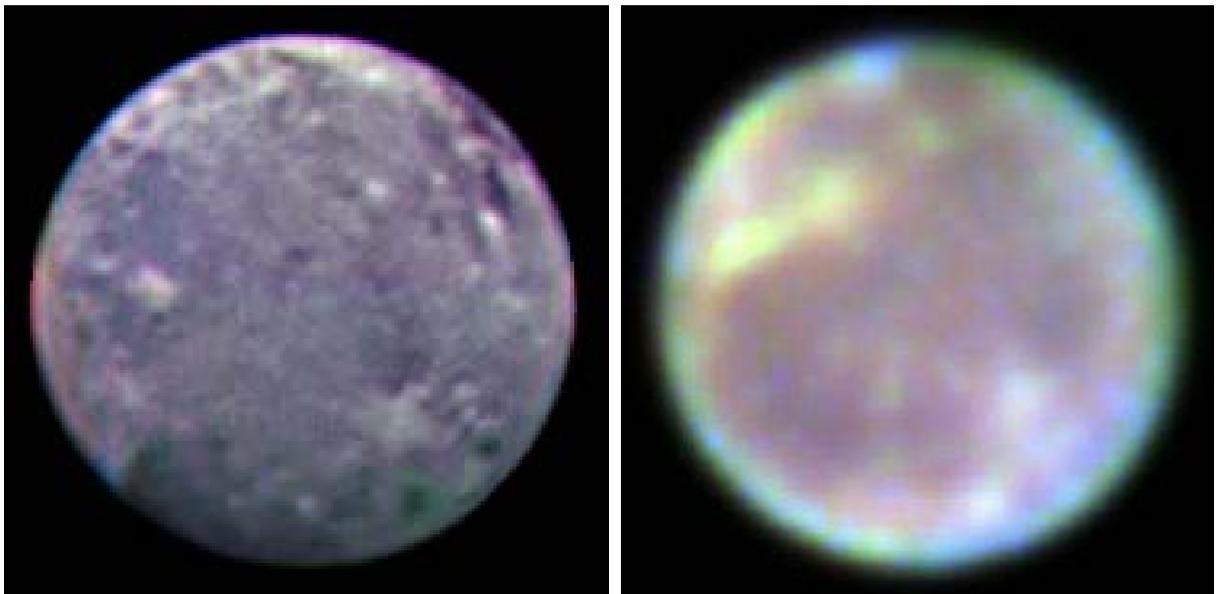

Fig. 2. Images of Ganymede with an approximate angular diameter of 1.6 arc seconds (7.8 μrad). (Left) BRI false color image obtained with PALM-3000 and TMAS (courtesy S. Hildebrandt). The pixel sampling in this image is 10 milli arc seconds (50 nrad), corresponding to ~ 35 km on the surface of Ganymede at this distance. (Right) Hubble Space Telescope false-color image of Ganymede for comparison (NASA image).

The PHARO science camera combined with the vector-vortex coronagraph [19] has directly imaged the exoplanets orbiting HR8799, as well as the debris disk around the star HD141569, showing the unprecedented inner working angle of this coronagraphic instrument [20]. The debris disk images show a clearing within the inner ring to 20AU along the projected semi-major axis, an impressive demonstration of the PALM-3000 correction coupled with the coronagraph's high contrast sensitivity.

### 3. PROPOSED NEW ARCHITECTURE

The conceptual architecture for PULSE is one that combines both the use of a laser guide star and a modest temporal and spatial order wavefront sensor trained on a natural guide star. One of the main benefits of a traditional laser adaptive optics system, whereby the laser and associated sensor provide the high-order correction, and a tip-tilt-focus sensor trained on a natural guide star, provides much greater sky coverage, as one only needs to be in the proximity of an object to provide the low-order natural guide star signal. Because of the larger subapertures and slower update rate required, one can use much fainter stars, which are much more numerous than brighter stars. The main drawback to using a single laser guide star is that of focal anisoplanatism, also known as the cone-effect, whereby the volume of atmosphere probed by the laser beacon is a cone, rather than the cylindrical volume of atmosphere that the light from astrophysical objects pass through. With a three-dimensional distribution of atmospheric turbulence, this places a fundamental limit on the performance of a single laser adaptive optics system. While additional lasers can be used to fill in the missing information [21, 22], there are additional tomographic reconstruction errors that ultimately limit performance to outside of the realm necessary for exoplanet adaptive optics.

Single natural guide star systems on the other hand do not suffer from the cone effect, but are limited by the brightness of the natural guide star source. As one tries to use fainter and fainter guide stars, the spatial and temporal sampling of the wavefront sensor which gives optimal adaptive optics performance becomes coarser and coarser [11]. Ultimately, as shown by Fig. 3, wavefront errors of ~100 nm RMS can be achieved only with stars of magnitude ~7 or brighter, in median or better seeing (curves b and c) and with high spatial sampling of the wavefront as indicated by the s64 wavefront sensor mode corresponding t0 8.2 cm subapertures.

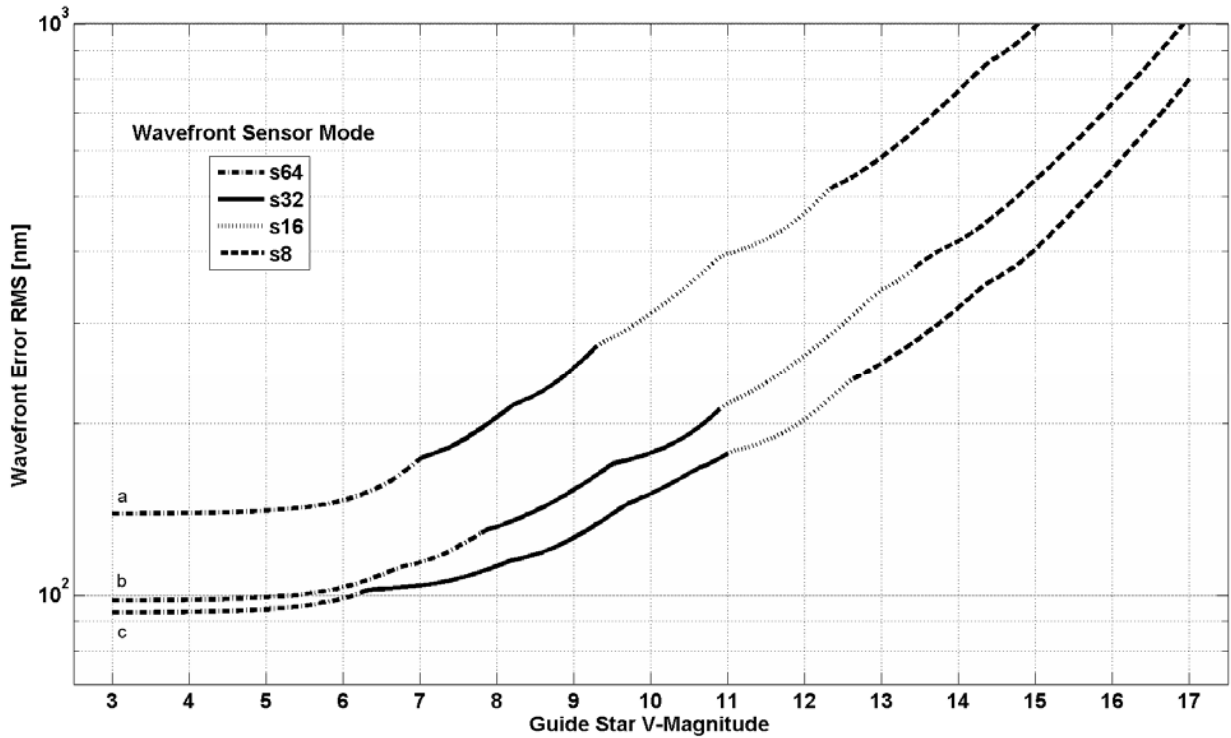

Fig. 3. Performance prediction for PALM-3000, from [1], with appropriate choice of optimal wavefront sensor mode and high-order wavefront sensor frame rate, under assumed conditions , a) $r_0$ = 6.0 cm, $r_0\text{eff}$ = 5.5 cm at $t_0\text{eff}$ = 0.79

ms b) $r_0$ = 9.2 cm, $r_0$eff = 8.4 cm, $t_0$eff = 2.46 ms, and c) $r_0$ = 15 cm, $r_0$eff = 13.8 cm, $t_0$eff = 2.46 ms, where the atmospheric parameters at zenith have been scaled to effective parameters corresponding to 30 degree zenith angle.

Fainter stars, in the magnitude range of 13 to 17, require much coarser sampling to deliver optimized performance, as indicated by the s8 wavefront sensor mode, corresponding to 63.7 cm subapertures, in Fig. 3.

The combination of a laser with a high-order sensor and a faint natural guide star with a modest-order natural guide star is proposed here to overcome the limitations of both individual systems as will be discussed in section 4. See Fig 4., the laser measurements will drive a high-order deformable mirror (3,388 active elements in the case of PALM-3000) in open-loop while an adaptive optics sharpened faint natural source will be measured by a low-order wavefront sensor, with commands sent in closed-loop to a second low-order deformable mirror. The natural guide star loop corrects for both the relatively weak low-order high-altitude turbulence as well as functioning as both the tip-tilt and low-bandwidth `truth' sensor loops in a traditional laser adaptive optics system.

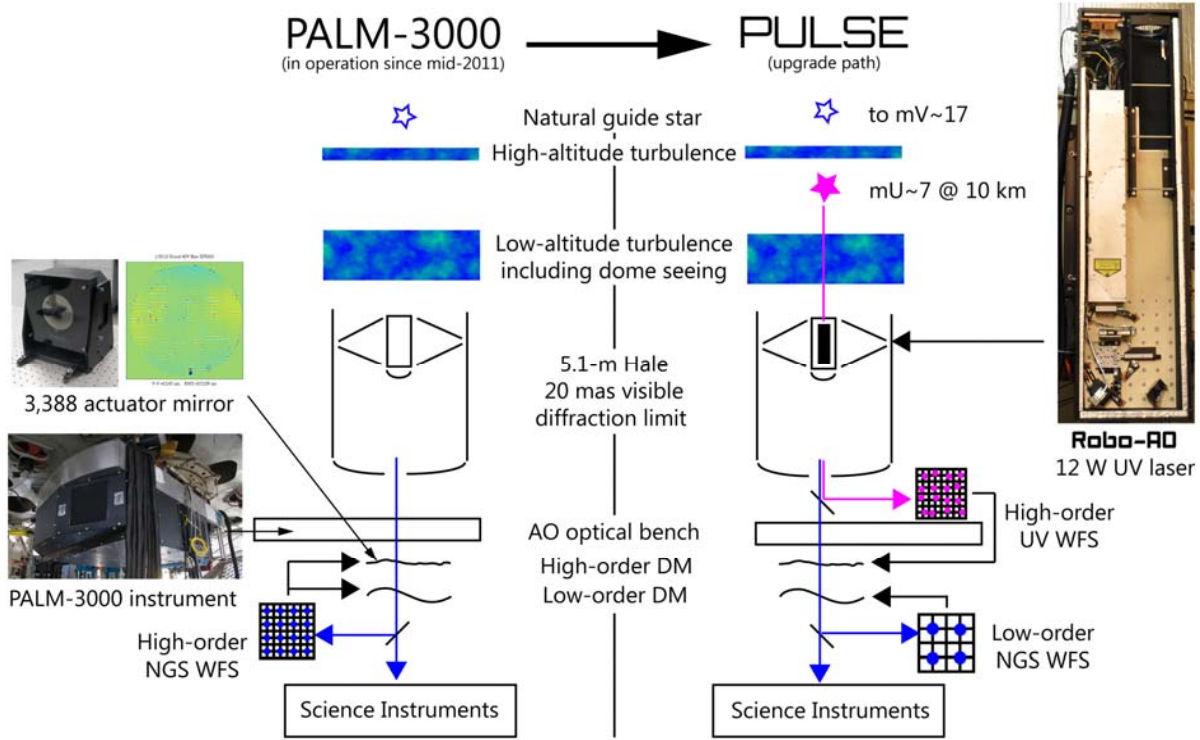

Fig. 4. Overview of proposed system architecture.

To upgrade the current PALM-3000 adaptive optics system, we will adopt the Robo-AO ultraviolet laser guide star [23, 24] (as opposed to using a 90-km Sodium laser guide star). Rayleigh beacons are preferred in this case because of their reduced cost (~USD10K/W), compact laser and projector assembly, ease of operation (turnkey system), fixed beacon distance (less complexity) and its classification as a class-1 laser with respect to overflying aircraft, so does not require human spotters normally required by the Federal Aviation Authority for visible lasers, greatly outweigh the additional focal anisoplanatism due to the lower, ~10km, beacon height. The Robo-AO ultraviolet laser guide star was first tested in September 2010 and was installed on a long-term basis in January 2011. As expected, the projected laser spot size has been consistently measured to be less than $\sqrt{2}$ times the visible seeing conditions. The measured photoreturn from a 375 m range gate at a distance of 10 km, is ~165 e-/subaperture/exposure at 1.2 kHz at zenith, which matches the theoretical return expected from the Lidar equation [25]. While we must coordinate propagation of the laser with US Strategic Command to avoid illuminating critical space assets, as with all laser AO systems, Robo-AO uses a newly developed system for laser deconfliction that opens the entire area above 50 degrees zenith distance for observation by requesting predictive avoidance authorization for ~700 individual fixed azimuth and elevation boxes of ~6 square degrees every night. This gives Robo-AO the capability to undertake laser observations of any target overhead at almost any time, increasing observing efficiency by removing the need to preselect targets of observation.

## 4. ADAPTIVE OPTICS PERFORMANCE SIMULATIONS

Here we analyze the performance of the proposed PULSE adaptive optics system on the Palomar 5-m telescope. The system uses a Rayleigh laser guide star to make high-order wavefront measurements. In addition, a natural guide star in the vicinity of the science field is used to make a low-order measurement. Traditionally, the NGS is used to make tip-tilt measurements only, since these cannot be measured using an LGS. We investigate here what the performance would be if the NGS sensor measures other low-order modes as well.

In order to evaluate the performance, simulations were run using the YAO Monte-Carlo simulation [26] tool. The wavefront sensor was modeled as an ideal slope sensor. This measures the mean slope over the subaperture directly from the wavefront and is much faster than simulating a physical optics model of the Shack-Hartmann wavefront sensor. It was verified that the two sets of simulations produce the same results for the on-axis, noiseless case.

The number of subapertures in the LGS wavefront sensor was assumed to be 40x40, which is slightly less than the highest 64x64 highest order that can be corrected by the DM. The NGS wavefront sensor has 5x5 subapertures. The LGS was assumed to measure all modes except for tip-tilt. The low-order modes were sent to a fictitious low-order mirror, while the high-order modes were sent to the real high-order DM. The measurements from the NGS were sent to a low-order mirror identical to the one for the LGS. Atmospheric parameters are tabulated in Table 1. The turbulence was assumed to be Kolmogorov (i.e., with an infinite outer scale) and an $r_0$ = 0.089 m at 500 nm. All the observations were carried out at zenith.

Table 1: Atmospheric parameters used in the simulations.

| Altitude (m) | 25 | 500 | 1000 | 2000 | 4000 | 8000 | 16000 |
|---|---|---|---|---|---|---|---|
| Turbulence Fraction | 0.667 | 0.128 | 0.016 | 0.036 | 0.059 | 0.058 | 0.036 |
| Layer speed (m/s) | 10 | 10 | 10 | 10 | 10 | 10 | 10 |
| Wind direction (deg.) | 0 | 76 | 92 | 190 | 255 | 270 | 350 |

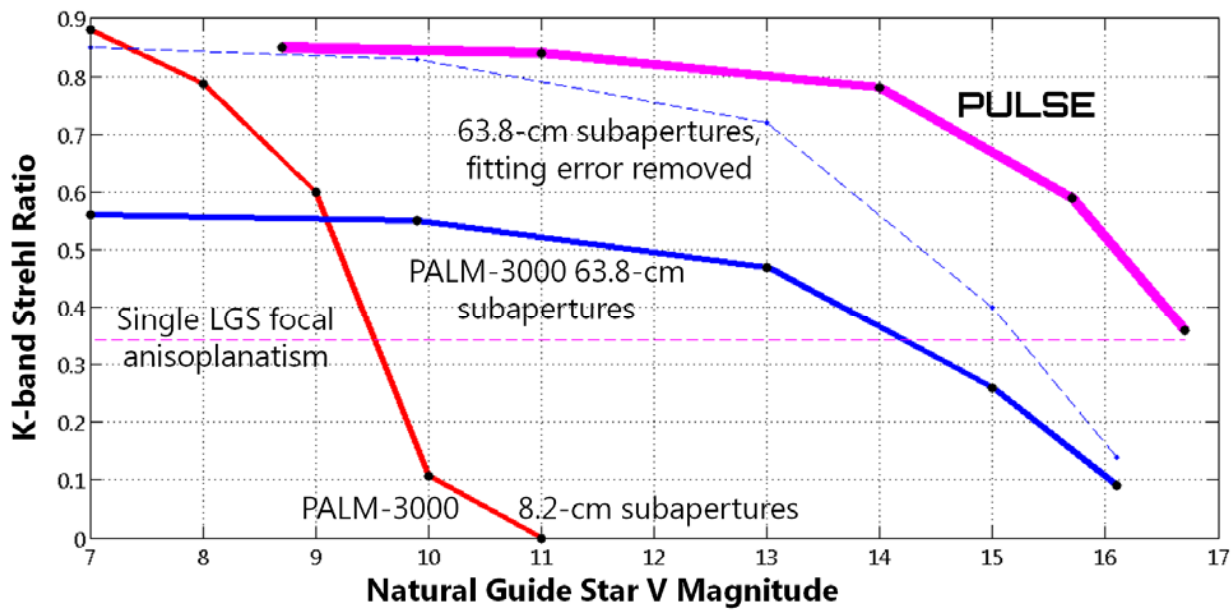

Fig. 5. K-band Strehl ratios for PALM-3000 and PULSE for different magnitudes of natural guide star. Five different curves are shown in the plot: (red) using a single natural guide star with the PALM-3000 s64 wavefront

sensor mode (corresponding to 8.2 cm subpaertures at the telescope aperture); (blue) using a single natural guide star with the PALM-3000 s8 wavefront sensor mode (corresponding to 63.8 cm subpaertures at the telescope aperture); (blue-dashed) using a single natural guide star with the PALM-3000 s8 wavefront sensor mode, but with the effects of fitting error artificially removed; (magenta-dashed) the limit of performance when using just a traditional single Rayleigh laser adaptive optics system; and (magenta) using the combination of a Rayleigh laser with a wavefront sensor with 12.8 cm subapertures (corresponding to 40 samples across the telescope aperture) and a natural guide star wavefront sensor with 1 m subapertures (corresponding to 5 samples across the telescope aperture).

In Fig. 5 we present the simulation results. On-axis Strehl ratios at K-band (2.2 microns) are presented. The red curve shows the expected Strehl ratio when using the highest spatial sampling of the PALM-3000 natural guide star wavefront sensor alone. Fainter than magnitude 7 stars, measurement error of the individual slopes begins to dominate the total error budget. The blue curve shows the PALM-3000 operating with its current lowest spatial sampling mode with a natural guide star. Performance is dominated by fitting error for stars brighter than ~14$^{th}$ magnitude (as also seen by the dashed blue line which shows the same system operating artificially in the absence of fitting error), and only for fainter stars do measurement and temporal errors start to add to the total wavefront error. Also shown in dashed magenta for reference is the limit imposed by focal anisoplanatism when using just a single Rayleigh laser guide star. The simulated PULSE performance is shown in magenta. While it is slightly lower in performance than PALM-3000 on bright stars (~85% Strehl vs. ~ 90% Strehl), the drop off with increasing stellar magnitude is much more gradual. For stars brighter than 11$^{th}$ magnitude, the high-order laser wavefront sensing and correction is compensating for the fitting error present that would otherwise be present if only a low-order natural guide star system were being used. For fainter stars, the PULSE performance drops off in a similar way to the low-order natural guide star only system as measurement and temporal errors start to increase with reduced flux. Note that the performance at 17$^{th}$ magnitude and beyond should asymptote to the single laser beacon focal anisoplanatism error line (dashed magenta) in the absence of tip-tilt errors.

## 5. DISCUSSION

The simulations presented here show a way to extend the exquisite adaptive optics correction of the just recently demonstrated extreme adaptive optics systems to much fainter guide sources than previously thought possible. This will enable for the first time direct imaging and characterization of exoplanets around the more numerous, cool and low-mass M-dwarf type stars. While the actual implementation is outside of the scope of this study, a conceptual way to implement this in the existing PALM-3000 system at the 5.1-m Palomar telescope has been presented.

We expect that the PULSE architecture presented here can be extended to other similar system. For example, the Keck II laser system is limited primarily by focal anisoplanatism, although is recently benefitting from a more powerful laser, an on-axis laser projector and a wide-field infrared tip-tilt sensor. By taking advantage of recent advances in detector technology, namely recently demonstrated small-format infrared avalanche photodiode arrays, one could augment the Keck II laser system with a low-order infrared pyramid wavefront sensor (which benefits from full-aperture image sharpening, and is therefore much more sensitive than a Shack-Hartmann sensor in this instance) to overcome the focal anisoplanatim in a way similar to that of the PULSE low-order natural guide star wavefront sensor.